\documentclass[twocolumn,showpacs,amsmath,amssymb,prd,nofootinbib]{revtex4}
\usepackage{epsfig}
\def\sss{\scriptscriptstyle}
\def\^#1{^{\sss #1}}
\def\_#1{_{\sss #1}}
\def\beq{\begin{equation}}
\def\eeqno#1{\label{#1}\end{equation}}
\def\ten#1#2{^{\sss#1}_{\sss#2}}

\def\az{a\_{0}}

\def\l0{\ell\_{0}}
\def\lm{\ell\_{M}}

\def\rar{\rightarrow}
\def\s{\sigma}
\def\l{\lambda}

\def\f{\phi}

\def\k{\kappa}

\def\r{\rho}

\def\m{\mu}
\def\n{\nu}
\def\e{\eta}
\def\Up{\Upsilon}
\def\C{\Gamma}
\def\A{\mathcal{A}}

\def\M{\mathcal{M}}

\def\Th{\hat\mathcal{T}}

\def\D{\Delta}
\def\Df{\D \f}
\def\d{\delta}

\def\a{\alpha}
\def\b{\beta}
\def\c{\gamma}
\def\d{\delta}
\def\eps{\epsilon}
\def\vr{{\bf r}}

\def\S{\Sigma}

\def\grad{\vec\nabla}

\def\gf{\grad\phi}

\def\RM{r\_M}

\def\gmn{g\_{\m\n}}
\def\Gmn{g\^{\mu \nu}}

\def\hGmn{\hat g\^{\mu \nu}}

\def\hmn{h\_{\m\n}}

\def\hab{h\_{\alpha\beta}}

\def\hlr{h\_{\lambda\rho}}

\def\hhlr{\hat h\_{\lambda\rho}}
\def\hgmn{\hat g\_{\m\n}}
\def\hhmn{\hat h\_{\m\n}}

\def\hgh{\hat g^{1/2}}
\def\gh{g^{1/2}}

\def\fh{\hat\f}
\def\gfh{\grad\fh}

\def\hC{\hat\C}

\def\T{\mathcal{T}}
\def\Tmn{\T\_{\m\n}}
\def\Th{\hat{\mathcal{T}}}
\def\hTmn{\Th\_{\m\n}}
\def\emn{\e\_{\m\n}}
\def\Emn{\e\^{\mu \nu}}

\def\azg{\A_0}
\begin{document}
\title{Gravitational waves in bimetric MOND}
\author{Mordehai Milgrom }
\affiliation{Department of Particle Physics and Astrophysics, Weizmann Institute}

\begin{abstract}
I consider the weak-field limit (WFL) of bimetric MOND (BIMOND) -- the lowest order in the small departures $\hmn=\gmn-\emn,~\hhmn=\hgmn-\emn$ from double Minkowski space-time. In particular, I look at propagating solutions, for a favorite subclass of BIMOND. The WFL splits into two sectors for two linear combinations, $\hmn\^{\pm}$, of $\hmn$ and $\hhmn$. The $\hmn\^+$ sector is equivalent to the WFL of general relativity (GR), with its gauge freedom, and has the same vacuum gravitational waves. The $\hmn\^-$ sector is fully nonlinear even for the weakest $\hmn\^-$, and inherits none of the coordinate gauge freedom. The equations of motion are scale invariant in the deep-MOND limit of purely gravitational systems. In these last two regards, the BIMOND WFL is greatly different from that of other bimetric theories studied to date. Despite the strong nonlinearity, an arbitrary pair of harmonic GR wave packets of $\hmn$ and $\hhmn$, moving in the same direction, is a solution of the (vacuum) BIMOND WFL.

\end{abstract}
\maketitle

\section{\label{introduction} INTRODUCTION}
MOND \cite{milgrom83} is a paradigm that replaces Newtonian dynamics and general relativity (GR); its goal is to account for all the mass discrepancies in the Universe without invoking dark matter (and ``dark energy''). Reference \cite{fm12} is an extensive recent review of MOND.
\par
Bimetric MOND (BIMOND) \cite{milgrom09} is one of several relativistic formulations of MOND (see Ref. \cite{fm12}). Aspects of it that have been considered so far are: matter-twin-matter interactions \cite{milgrom10b}, matter fluctuations in cosmology \cite{milgrom10c}, and aspects of cosmology \cite{cz10}. Still, BIMOND remains largely unexplored, despite its promise in several regards: It tends to GR for $\az\rar 0$ ($\az$ is the MOND constant); it has a simple and elegant nonrelativistic (NR) limit; it describes gravitational lensing correctly; and, it has a generic appearance of a cosmological constant that is of order $\az^2/c^4$ -- a well-known, observed coincidence.
\par
In particular, the important issue of propagating gravitational waves (GWs) has not been addressed. Here I begin to consider this question. Of the various aspects of GWs, such as emission, propagation in vacuum, propagation on a background, and interaction with matter (and detection), I treat here mainly the simplest -- propagation on a doubly Minkowski space-time background. This is important as it concerns the speed of propagation and existence of superluminal or subluminal GWs, with the issue of causality. It also concerns stability and the possible appearance of ghost modes.
Aspects of emission and detection are touched on briefly in Sec. \ref{emission}.
Waves in another relativistic MOND theory, TeVeS \cite{bekenstein04} -- with a very different behavior than BIMOND -- have been studied in Refs. \cite{bekenstein04,sagi10}.
\par
In Sec. \ref{bimond}, I present BIMOND in brief, Sec. \ref{wfl} describes the weak-field limit, Sec. \ref{phenomen} discusses phenomenological constraints on the BIMOND interaction, Sec. \ref{gw} discusses GW propagation, and Sec. \ref{emission} discusses briefly emission and detection of GWs.
\section{\label{bimond} BIMOND IN BRIEF}
%{\it BIMOND IN BRIEF}--
In BIMOND, gravity is described by two metrics, $\gmn$ and $\hgmn$. Its action is:
 $$ I=-\frac{c^4}{16\pi G}\int[\b \gh R +\a\hgh \hat R
 +2v\_{g\hat g}\lm^{-2}\M]d^4x$$
\beq +I\_M(\gmn,\psi_i)+\hat I\_M(\hat g\_{\m\n},\chi_i).  \eeqno{gedat}
Here, $G$ is Newton's constant, $R$ and $\hat R$ are the Ricci scalars of $\gmn$ and $\hgmn$, whose determinants are $-g$ and $-\hat g$, and $v\_{g\hat g}$ is a combined volume form.\footnote{It makes sense to take $v\_{g\hat g}$ to be symmetric in the two metrics, and to reduce to $g^{1/2}$ when the metrics are equal; e.g., $v\_{g\hat g}=(g^{1/2}+\hat g^{1/2})/2$, or $v\_{g\hat g}=(g\hat g)^{1/4}$.}  The MOND length, $\lm\equiv c^2/\az$, naturally sets the strength of the coupling. The dimensionless interaction, $\M$, is a function of the  ``relative-acceleration'' tensors $\lm C\^\a\_{\b\c}$, where $C\^\a\_{\b\c}=\C\^\a\_{\b\c}-\hat\C\^\a\_{\b\c}$, and
$\C\^\a\_{\b\c}$,  $\hat\C\^\a\_{\b\c}$ are the Levi-Civita
connections.\footnote{$\M$ may also depend on scalars such as $\k\equiv(g/\hat g)^{1/4}$, or $\Gmn\hgmn$.} $I\_M$ and $\hat I\_M$ are the matter actions for matter and twin matter (TM) respectively. Hereafter I take $c=1$.
\par
BIMOND is required to reduce to GR in its decoupling limit
$|C\^\a\_{\b\c}|\gg\az$ (or $\az\rar 0$), where it has to reduce to two uncoupled copies of GR. MOND, in general, and BIMOND in particular, may well not be perturbatively expandable in powers of $\az$ near  $\az=0$.\footnote{For example if the MOND interpolating function behaves as ${\rm tanh}(a/\az)$.} The assumption that such an expansion is possible strongly underlies the damning verdict of Ref. \cite{boulanger01} regarding multimetric theories.
BIMOND, thus, may well escape this verdict.

\section{\label{wfl} THE WEAK-FIELD LIMIT}
Consider the weak-field limit (WFL) in the context of a matter-TM symmetric universe, which makes the most sense for BIMOND \cite{milgrom10c}. Then, the homogeneous cosmological metrics are equal, and locally there is a coordinate frame in which they are both Minkowskian. The WFL pertains to small departures from this background:
 \beq \gmn=\emn+\hmn,~~~~~\hgmn=\emn+\hhmn, ~~~~~\hmn,~\hhmn\ll 1,  \eeqno{gusta}
where we take the lowest order in $\hmn,~\hhmn$.
\par
First, replace $R$ in Eq.(\ref{gedat}) by
\beq -\C^{(2)}\equiv -\Gmn(\C\ten{\c}{\m\l}\C\ten{\l}{\n\c}
-\C\ten{\c}{\m\n}\C\ten{\l}{\l\c}), \eeqno{nirta}
and, similarly, $\hat R$ by $-\hC^{(2)}$. This only adds a divergence to the Lagrangian density.
\par
Now replace $\C^{(2)}$ and $\hat\C^{(2)}$ by their weak-field expressions $\bar\C^{(2)}(\hmn)$ and $\bar\C^{(2)}(\hhmn)$. These are of order $(h\_{\m\n,\l})^2$ and $(\hat h\_{\m\n,\l})^2$. Close scrutiny shows that in $\M$ and its volume prefactor we can replace everywhere $\gmn$ and $\hgmn$ by $\emn$, $g\_{\m\n,\s}$ by  $h\_{\m\n,\s}$, and  $\hat g\_{\m\n,\s}$  by $\hat h\_{\m\n,\s}$.
\par
Note importantly that while the WFL is defined by $|\hmn|\ll 1$, $|\nabla h|/\az$, is not assumed small ($\nabla h$ stands schematically for $h\_{\m\n,\l},~ \hat h\_{\m\n,\l}$): In our approximation scheme, $\az$ is considered of order $\nabla h$. Seen differently, up to a cosmological-constant term of order $\az^2$, in the WFL we can write $\az^2\M$, schematically,
as $(\nabla h)^2 (\az/\nabla h)^2\{\M[(\nabla h/\az)^2]-\M(0)\}$. In MOND, $u^{-1}[\M(u)-\M(0)]$ is bounded by a number of order unity (see below). This justifies our handling of $\az^2\M$ in the WFL, as any correction that we leave out is of higher order in $\nabla h$.
\par
In our approximation, $C\^\a\_{\b\c}$ become linear in first derivatives of $ \hmn\^-\equiv \hmn-\hhmn$.
Thus, $\M$ becomes a functional, $\M^*(q\^-)$, of $\hmn\^-$ through $q\^-$, which stands for variables of the form $(h\^-\_{\m\n,\s}/\az)^2$. They are of zeroth order in our approximation (and so are all appearances of $\M$ in the WFL).
\par
Also,
$$I\_M\approx \frac{1}{2}\int \hmn\T^{\m\n}d^4x,~~~\hat I\_M\approx \frac{1}{2}\int \hhmn\hat\T^{\m\n}d^4x,$$
where $\T^{\m\n},~\hat\T^{\m\n}$ are the Minkowskian energy-momentum tensors.
Assuming $\a+\b\not=0$ (the case $\a+\b=0$ requires special treatment), define:
\beq \bar\M(q\^-)\equiv \M^*(q\^-)-\frac{\a\b}{2\az^2(\a+\b)}\bar \C^{(2)}(\hmn^-). \eeqno{lopat}
[Equation (\ref{lopat}) is all zeroth order.] Putting all this together, we have
 $$ (\a+\b)I\approx -\frac{1}{16\pi G}\int [-\bar \C^{(2)}(\hmn^+)+8\pi G\hmn^+\T^{+\m\n}$$
\beq +2(\a+\b)\az^2\bar\M(q\^-)
+8\pi G\hmn^-\T^{-\m\n}]d^4x,  \eeqno{gepada}
where $\hmn^+\equiv \b\hmn+\a\hhmn$,
$\T^{+\m\n}\equiv \T^{\m\n}+\hat\T^{\m\n}$, and $\T^{-\m\n}\equiv\a\T^{\m\n}-\b\hat\T^{\m\n}$.
\par
So, the WFL {\it gravitational} action reduces to two decoupled terms for $\hmn^+$ and $\hmn^-$. The field equations for $\hmn^+$ are identical with the WFL of GR, but are sourced by $\T^{+\m\n}$.
The equations for the $\hmn^-$ sector are sourced by $\T^{-\m\n}$, and carry the MOND modification.
\par
The gauge freedom is afforded by invariance under $x^\m\rar x'^\m=x^\m+\eps^\m(x)$, with $\eps_{\m,\n}\ll 1$ of the same order as $\hmn,~\hhmn$. These transform $\hmn\rar\hmn-\eps_{\m,\n}-\eps_{\n,\m}$, $\hhmn\rar\hhmn-\eps_{\m,\n}-\eps_{\n,\m}$. So $\hmn^-$ is not affected, while $\hmn^+\rar\hmn^+ -(\a+\b)(\eps_{\m,\n}+\eps_{\n,\m})$. So, the $\hmn^+$ sector enjoys the same gauge freedom as in the WFL of GR, while the $\hmn^-$ sector is not subject to the above coordinate gauge freedom.
\par
A particularly apt choice \cite{milgrom09} of the scalar arguments of $\M$ are $\Up\equiv  \Gmn\Up\_{\m\n}$, and $\hat\Up\equiv  \hGmn\Up\_{\m\n}$, where
\beq\Up\_{\m\n}=C\ten{\c}{\m\l}C\ten{\l}{\n\c}
-C\ten{\c}{\m\n}C\ten{\l}{\l\c}. \eeqno{mulpat}
These scalars have the same structure as $\C\^{(2)},~\hC\^{(2)}$. This choice leads to a particularly simple NR limit, and, as we shall see, endows this version of BIMOND with a certain welcome affinity to GR.
\par
In the WFL, $\Up\approx\hat\Up\approx \bar\Up\equiv\bar\C^{(2)}(\hmn\^-)$. Thus, for this choice of variables, $\bar\M$ depends on $h\^-\_{\m\n,\c}$ through a single variable, $z\equiv-\bar\Up/2\az^2$;
\beq z=\frac{1}{8\az^2}[{{h\^{-\n\r}}\_{,}}\^\c(h\^-\_{\n\r,\c}-2h\^-\_{\n\c,\r})
-{{h\^{-}}\_{,}}\^\c(h\^-\_{,\c}-2h\^{-\r}\_{\c,\r})].  \eeqno{kiolio}
(Indices are raised and lowered with $\emn$ in the WFL.)
\par
The field equations for $\hmn\^-$ then read
\beq \S\_{\m\n}\equiv[\bar\M'(z)\bar S\ten{\l}{\m\n}]\_{,\l}
=\frac{8\pi G}{\a+\b}\T^{-}_{\m\n},  \eeqno{biupo}
where
\beq  \bar S\ten{\l}{\m\n}\equiv C\ten{\l}{\m\n}-\frac{1}{2}\d\^{\l}\_{\m}C\_{\n}
-\frac{1}{2}\d\^{\l}\_{\n}C\_{\m}+\frac{1}{2}\emn(C\^{\l}-\bar C\^{\l}), \eeqno{cucupo}
and in the WFL, $ C\^\a\_{\b\c}\approx
\frac{1}{2}\e\^{\a\s}(h\^-\_{\b\s,\c}
+h\^-\_{\c\s,\b}-h\^-\_{\b\c,\s})$.
 [$C\_{\n}=(1/2)h\^{\l -}\_{\l,\n}$, $C\^{\m}\equiv \Emn C\_{\n}$].
\par
$-\bar S\ten{\l}{\m\n,\l}=
\bar G\_{\m\n}(\hab\^-)$ is the WFL of the Einstein tensor for $\emn+\hmn\^-$; it satisfies the Bianchi identities $\bar S\ten{\l}{\m\n,\l,}{\^\n}=0$.
\par
The equations of the $\hmn\^+$ sector satisfy the usual four Bianchi identities.
Since
$\S\_{\m\n}$ do not satisfy Bianchi identities (if $\bar\M'$ is not constant), Eq.(\ref{biupo}) constitutes ten independent equations for the components of $\hmn\^-$.
\par
In both the $\hmn\^{\pm}$ sectors there is also a cosmological constant of order $\az^2\M(0)$, which I neglect, consistent with our approximation (in the Lagrangian such a term is of order $h\az^2$).
\par
For the special choice of parameters $\a+\b=0$ (and taking $\b=1$ for concreteness),
the field equations can be written as
\beq \bar G\_{\m\n}(\hlr^-)=-8\pi G (\Tmn+\hTmn),  \eeqno{milka}
 \beq \bar G\_{\m\n}(\hlr)=\S\_{\m\n}(\hlr^-)-8\pi G \Tmn.  \eeqno{lipka}
Equation (\ref{milka}) is identically divergenceless, but $\hmn\^-$ has no gauge freedom. Equation (\ref{lipka}) implies the 4 equations $\S\^{\m\n}\_{,\m}(\hlr^-)=0$, which together with the six independent equations in Eq.(\ref{milka}) give a well-defined Cauchy problem for $\hmn\^-$. After $\hmn\^-$ is solved for, it is substituted into Eq.(\ref{lipka}), which gives exactly the WFL of GR for $\hmn$ (which has all the gauge freedom, and to which matter couples) only with the extra source $-(8\pi G )^{-1}\S\_{\m\n}(\hlr^-)$, which plays the role of phantom (``dark'') matter for the $\hmn$ (matter) sector.
MOND phenomenology in the NR limit of this theory \cite{milgrom10a,milgrom10b} requires here
 $\bar\M'(z\rar\infty)\rar 0$, and $\bar\M'(z\ll 1)\approx -z^{-1/4}$.
\section{\label{phenomen} PHENOMENOLOGICAL CONSTRAINTS ON THE FORM OF THE INTERACTION}
To further investigate the $\hmn^-$ sector in the WFL we need to know $\bar\M(z)$, which is constrained by MOND phenomenology in the NR limit. This limit is a special case of the WFL where, further to the smallness of $\hmn$ and $\hhmn$, the metrics are time independent.
It was thoroughly investigated in Refs. \cite{milgrom09,milgrom10a,milgrom10b} for the subclass of BIMOND with the variables $\Up,~\hat\Up$, which proved to be particularly felicitous.\footnote{Reference \cite{milgrom09} also explored a little the more general case.} It was shown that there is a choice of gauge for which  $\hmn=-2\f\d\_{\m\n}$, $\hhmn=-2\hat\f\d\_{\m\n}$ (as in GR), in which case
\beq z=(\gf-\gfh)^2/\az^2. \eeqno{mjui}
{\it NR MOND phenomenology thus probes only non-negative values of the argument of $\bar\M$.}
\par
The low-acceleration, or deep-MOND limit (DML) is defined by applying a space-time scaling transformation [$(t,\vr)\rar\l(t,\vr)$], with the scale factor $\l\rar\infty$ (so accelerations $a\rar\l^{-1}a\rar 0$). Equivalently, we can always normalize the various degrees of freedom such that the limit is attained by taking in the theory $\az\rar\infty$, $G\rar 0$, with $\azg\equiv G\az$ finite \cite{milgrom14}.
If such a limit exists for some theory, it is automatically scale invariant (SI). This is a defining property of the MOND paradigm, and yields MOND phenomenology (for example, asymptotic flatness of rotation curves) \cite{milgrom09a}.

The limit also has to be nontrivial, i.e., retain $\azg$ as a pivotal dimensioned constant. (For example, the limit for Newtonian dynamics exists, but gives instead of the Poisson equation $\Df=0$, which is uninteresting, or trivial.) These require for the case $\a+\b\not =0$ that
\beq \bar\M'(z\ll 1)\propto z^{1/2}.   \eeqno{mbaloa}
[With a power $<1/2$ the theory becomes trivial ($\azg$ disappears), with a power $>1/2$ the limit does not exist.]
\par
This behavior has been verified down to $z\sim 10^{-3}$ (e.g., Ref. \cite{milgrom13}); and, one usually assumes it down to $z=0$.
\par
In this case, the field equations (\ref{biupo}) for the $\hmn^-$ sector -- now reading $[(-\bar\Up)^{1/2}\bar S\ten{\l}{\m\n}]\_{,\l}
\propto\azg\T^{-}_{\m\n}$ -- are nonlinear down to zero field. Furthermore, they are SI for purely gravitational systems.\footnote{The action is not SI, but is multiplied by a constant under scaling.}
In these regards, the WFL of BIMOND is very different from that of other bimetric theories studied to date.
\par
Also, the equations for $\hmn^+$ become the vacuum Einstein equations, $\bar G\_{\m\n}(\hab\^+)=0$ (since in the DML,  $G\rar 0$), which are also SI, as are the geodesic equations and the gauge conditions, which do not involved dimensioned constants. So, not only the DML of NR BIMOND is SI, but also, more generally, the DML of its WFL: given a solution of the theory with metrics $\hmn^\pm(x)$, and particles $i$ having world lines $x_i^\m(q)$, the metrics $\hmn^\pm(x/\l)$, with world lines $\l x_i^\m(q)$ are also a solution (with initial and boundary conditions transformed correspondingly).
\par
It follows, for example, that the light-bending angle is independent of the impact parameter, $b$, for a static lensing mass $M$, well within $b$ (so it can be considered a point mass), and $b\gg \RM\equiv (MG/\az)^{1/2}$ ($\RM$ is the MOND radius of the mass), so that the DML applies.
\par
The order of taking the WFL and the deep-MOND ($z\rar 0$) limit (DML) is of consequence. Scrutiny shows that it is proper to take the WFL first: the MOND length, $\lm$, is larger than the Hubble distance. If the metric(s) vary on scale $L$, the DML corresponds to $\hmn/L\ll \lm^{-1}$. So, in all sub-Hubble applications, where $L\ll\lm$, the MOND small parameter is much larger than the relativity small parameter: $\hmn/L\az\gg\hmn$.
\par
The WFL theory with $\a=-\b$ [Eqs.(\ref{milka}
-\ref{lipka})] also has a formal, nontrivial DML, which is thus SI: Define $\hlr^*\equiv \az\hlr^-$, taken to scale as $\hlr^*\rar\l^{-1}\hlr^*$, to match its dimensions. Then in the DML, $\bar G\_{\m\n}(\hlr^*)=-8\pi\azg (\Tmn+\hTmn)$,  with $\S\^{\m\n}\_{,\m}(\hlr^*)=0$, [in the DML, $\S$ scales as $d(dh^*/dx)^{1/2}/dx$]. And, $\hlr=\hhlr$,\footnote{So the initial and boundary conditions have to satisfy this.} satisfy
$\bar G\_{\m\n}(\hlr)=\S\_{\m\n}(\hlr^*)$ (and SI gauge conditions), which involves no constants.
\par
It is an interesting, open question whether BIMOND, not restricted to the WFL, has a nontrivial DML.
\par
In the decoupling limit, $z\rar\infty$ ($\az\rar 0$), the NR theory has to approach Newtonian dynamics. This implies that $\bar\M'(z)$ has to approach a certain constant value $a(\a,\b)$ there. For $\b=1$ ($\a\not= -1$), $a(\a,1)=-\a/(1+\a)$. This ensures also that the fully relativistic BIMOND, for the $\gmn$ sector, tends to GR in this limit. Such GR compatibility, which can be made as fast as desired with an appropriate form of $\M$, is required for BIMOND to comply with Solar-System and binary-pulsar limits. For $\b\not= 1$ it is not known whether the condition for Newtonian compatibility also ensures GR compatibility. If it does not, we would have a good reason to prefer $\b=1$, in addition to other reasons \cite{milgrom09}.
\par
As mentioned already, $\bar\M'(z)$ may not be expandable in powers of $\az$ near $\az=0$. If true, this would hopefully obviate the obstacles raised in Ref. \cite{boulanger01} regarding general multimetric theories. But this is yet to be checked.

\section{\label{gw} GRAVITATIONAL WAVES}
A general comment is in order here regarding the characteristic accelerations of GWs in comparison with $\az$.
Away from sources, GWs are weak in the relativity sense ($|\hmn|\ll 1$). However, they need not be so in the MOND sense.
In the WFL of BIMOND, the relevant measure of this is $a/\az\sim\lm|\nabla\hmn\^-|$. For wavelength $\l$, and amplitude $h$, $a/\az\sim\lm h/(\l/4)\approx (8\pi D_H/\l)h\approx 10^{19}h(\n/{\rm Hz})$, where $D_H\approx\lm/2\pi$ is the Hubble distance. This can be much larger than 1 even very far from sources. Write the (Einsteinian) amplitude of GWs produced by a source of mass $M$ at a distance $D$ as
$h= fMG/D$, where $f$ is a numerical factor that depends on details of the emitter: how relativistic it is, the fraction of the ``nonspherical mass'', geometry, etc. Also, write for the typical wavelength $\l=qR_s$, where $R_s=2MG/c^2$. Then $a/\az\sim 4\pi fD_H/qD$. So, for $f/q\not\ll 1$ we can have $a/\az>1$ even to the Hubble distance.\footnote{Because of the inherent nonlinearity of MOND we cannot consider separately the typical ``acceleration'' of different frequencies as done here. So this estimate is only a rough indication.} This fact rather complicates the analysis of GW in MOND. For example, it means that the ``accelerations'' produced by the GW itself may sometimes be comparable, or even much larger than those produced by the background on which the GW is moving.
\par
In the WFL of BIMOND, vacuum gravitational waves are solutions of the equations
\beq \bar G\_{\m\n}(h\^+)=0,~~~~~~ \S\_{\m\n}=[\bar\M'(-\bar\Up/2\az^2)\bar S\ten{\l}{\m\n}]\_{,\l}=0. \eeqno{biuposht}
For $\hmn^+$, which satisfy a vacuum Einstein equation, we may take the standard harmonic gauge, and we get for them the two standard modes of GR. The equation for $\hmn\^-$ is nonlinear even for the smallest of amplitudes.
\par
Interestingly, despite the strong nonlinearity, an arbitrary Einsteinian, plane-wave packet,\footnote{Namely, an arbitrary superposition of Einsteinian plane waves with various momenta, all in the same direction.} $\hmn\^-$, satisfying the harmonic gauge, is a solution of the second field equation (\ref{biuposht}).\footnote{No gauge freedom for $\hmn\^-$ is employed; we simply find that an Einsteinian packet that satisfies the harmonic gauge, whose momenta are thus all null, is a solution.}
This follows since such a packet annuls $\bar\Up$, the argument of the WFL interaction.
To see this, consider the momentum-space representation. In $k$ space, all the scalars, such as $\bar\Up$, are sums of terms containing  $k'\_\a\eps\_{\b\c}(k')k\_\m\eps\_{\n\l}(k)$, with indices contracted in pairs by three $\e\^{\s\r}$; $\eps\_{\m\n}(k)$ is the $k$ component of $\hmn\^-$. The harmonic condition reads $k\^\m\eps\_\m\^\n(k)=(1/2)k\^\n\eps(k)$, where  $\eps\equiv \eps\^\m\_\m$. For vacuum solutions of the WFL Einstein equation, this implies $k^2=0$ for all $k$: all momenta are mutually proportional null vectors; so, for any two $k\^\m k'\_\m=0$, and $k'\_\m\eps\^\m\_{\l}(k)=
(1/2)k'\_\l\eps(k)$. Contracting all indices in the above expression must thus lead to the contraction of two momenta and gives zero. Seen differently: apply a longitudinal Lorentz boost under which the scalars, which are also Lorentz scalars, should not vary. The transverse components of $\eps\_{\m\n}(k)$ are unaffected, and the longitudinal ones, which can be written as $k\_\n e\_\m+k\_\m e\_\n$ for some $e\_\m$, do not contribute since they lead to products of three $k$s. But, all momenta are multiplied by the same Lorentz factor; so to be invariant the scalars must vanish. So even though rough scaling may imply high accelerations for such waves, in fact they annul the argument of $\bar\M$.
\par
A similar situation occurs in nonlinear theories of electromagnetism, such as the Born-Infeld theory, where Maxwellian plane-wave packets annul the electromagnetic invariants. There, however, the (Lorentz) invariants are also gauge invariant, while here the scalars are not invariant to changes of the longitudinal components: $\eps\_{\m\n}(k)\rar\eps\_{\m\n}(k)+k\_\n e\_\m(k)+k\_\m e\_\n(k)$.
\par
So, we can write in Eq. (\ref{biuposht}) $\bar\M'(0)\bar S\ten{\l}{\m\n,\l}=0$, and $\bar S\ten{\l}{\m\n,\l}(\hab\^-)=0$ is the WFL Einstein equation. So,
if $\bar\M'(z)$ does not diverge at $z=0$ [it vanishes at $z=0$, by Eq.(\ref{mbaloa})], two arbitrary Einsteinian wave packets for $\hmn\^+$ and $\hmn\^-$, satisfy Eqs.(\ref{biuposht}). This is true for the choice of $\Up,~\hat \Up$ as variables of $\M$, since it is for this case that we have $\bar S\ten{\l}{\m\n,\l}=0$. So, it is not clear that our packet is also a solution of BIMOND for a more general choice of variables of $\M$.
\par
Thus, a pair of arbitrary, Einsteinian, plane-wave, harmonic packets of $\hmn$ and $\hhmn$ {\it moving in the same direction}, is a vacuum solution of BIMOND, since they can be combined into such packets of $\hmn\^{\pm}$.

\section{\label{emission}EMISSION AND DETECTION}
The objects and phenomena we have in mind as sources of potentially detectable GWs are all characterized by accelerations much higher than $\az$.\footnote{Any relativistic system much smaller than the Hubble length is of high acceleration.} Thus, in versions of BIMOND that are compatible with GR, as we assume, GW emission occurs as in GR: only $\gmn$ waves are emitted by matter systems (and $\hgmn$ waves by TM systems), and they have GR characteristics. When leaving the high-acceleration region around the emitter, such waves would gradually (and in a manner yet to be studied) be transfigured into modes of well-defined propagation properties in that regime (for example $\hmn\^{\pm}$ modes). Thus for example, an asymptotic $\hmn\^+$ wave, which is a combination of $\gmn$ and $\hgmn$ waves can be produced by matter or TM sources.
\par
Foreseeable terrestrial and space detectors (but not pulsar timing arrays) are within the very-high-acceleration field of the inner Solar System. Any BIMOND GWs that enter this region are describable within the inner Solar System as combinations of $\hmn$ and $\hhmn$ waves, each with the two polarization modes as in GR, and with only the $\hmn$ waves coupling to the (matter) detectors. So BIMOND does not predict that GWs with polarizations other than those predicted by GR will be detected.
However the relation between the GWs that are detected and their sources are expected to be different than in GR. For example, both matter and TW sources may produce GWs that arrive in the inner Solar System as a combination containing $\gmn$ waves, which are detectable by matter detectors. This may lead to GWs without an electromagnetically detectable (TM) source. Also, propagation properties on backgrounds, which have not yet been studied, may lead to different propagation times for gravitational waves and light:
GW propagation may respond to both $\gmn$ and $\hgmn$ backgrounds, while photons are affected only by a $\gmn$ background.

%\clearpage
\end{document}